\documentclass[12pt]{iopart}

\usepackage{graphicx,color}
\usepackage{amssymb}
\usepackage{amsbsy}
\usepackage{chngcntr}
\usepackage{amsfonts}
\usepackage{verbatim}

\renewcommand{\(}{\left(}
\renewcommand{\)}{\right)}

\newcommand{\lk}{\left[}
\newcommand{\rk}{\right]}

\def\roughly#1{\mathrel{\raise.3ex
\hbox{$#1$\kern-.75em\lower1ex\hbox{$\sim$}}}}

\newcommand{\be}{\begin{equation}}
\newcommand{\ee}{\end{equation}}
\newcommand{\bqa}{\begin{eqnarray}}
\newcommand{\eqa}{\end{eqnarray}}

\begin{document}

\rightline{BI-TP 2014/17}
\rightline{TUW-14-13}

\title{Scalar field  collapse  with negative cosmological constant}

\author{ R.~Baier $^1$, H.~Nishimura $^1$ and  S.~A.~Stricker  $^2$  }

\address{$^1$Faculty of Physics, University of 
Bielefeld, D-33501 Bielefeld, Germany}
\address{$^2$Institute for Theoretical Physics, Vienna University of Technology, Wiedner Hauptstr. 8-10, A-1040 Vienna, Austria}

\ead{baier@physik.uni-bielefeld.de}
\ead{nishimura@physik.uni-bielefeld.de}
\ead{stricker@hep.itp.tuwien.ac.at}

\begin{abstract}
The formation of black holes or naked singularities is studied in a model in which a homogeneous time-dependent scalar field with an exponential potential couples to four dimensional gravity with negative cosmological constant.
 An analytic solution is derived and its consequences are discussed.
The model depends only on one free parameter, which determines the equation of state and decides the fate of the spacetime. 
Without fine tuning the value of this parameter the collapse  ends in a generic
formation of a black hole or a naked singularity. The latter case violates the cosmic censorship conjecture. 

\end{abstract}

\maketitle
\section{Introduction}

The aim of this work is to investigate the cosmic censorship conjecture
\cite{Penrose:1969pc} in more detail and its violation in a spacetime with negative cosmological constant 
motivated by the recent work of Hertog et al.
\cite{Hertog:2003zs,Hertog:2003xg,Hertog:2004gz}, 
Hubeny et al. \cite{Hubeny:2004cn,Rangamani:2004iw} and Dafermos \cite{Dafermos:2004ws}.
The cosmic censorship conjecture states that singularities are hidden behind an event horizon and therefore are not accessible for outside observers.
However, there are counterexamples, in particular in the scalar field model under consideration, in which naked singularities
\cite{Joshi:2002,Singh:1997wa,Joshi:2011hb,Joshi:2012mk,Joshi:2009zza} 
are produced without fine-tuning initial conditions.
For a different situation of scalar field collapse see \cite{Gundlach:1999cu} for a review.
Here we extend the models of gravitational  scalar field collapse in general relativity with
vanishing and  non-vanishing cosmological constant, which are widely studied in the literature,
for example in \cite{Joshi:2008zz,Goswami:2006ph,Goswami:2007na,Joshi:2007zza,Husain:1994uj}
and \cite{Cissoko:1998mx,Deshingkar:2000hd,Madhav:2005kg}, respectively.

 The formation of singularities is  particularly interesting in the context of the AdS/CFT correspondence \cite{Maldacena:1997re} where type IIB string theory on asymptotically $AdS_5 \times S_5$ is conjectured to be  dual to  $\mathcal{N}=4$ Super Yang Mills theory (SYM) living on the boundary of $AdS_5$.
Taking the correspondence at face value, we expect that the field theory, which is well behaved over the entire evolution, should be able to resolve the (naked) singularities on the gravity side and therefore be able to provide some insights into quantum gravity. For this reason it is important to find examples where naked singularities form.

In the supergravity approximation, which is dual to a strongly coupled field theory, 
analytic examples of 
 black hole formation \cite{Baier:2012tc,Steineder:2012si,Baier:2013gsa,Stricker:2013lma,Stricker:2014cma} are very important and useful to understand thermalization   in strongly coupled systems such as in ultra-relativistic heavy-ion collisions \cite{DeWolfe:2013cua,CasalderreySolana:2011us}  or cold atomic gases.
The interplay between black hole and naked singularity formation in  AdS
gravitational collapse still requires further investigations in case of  thermalization dynamics.

In section 2 the model and its analytic solution is described, including the limiting behaviour of vanishing cosmological constant $\Lambda \rightarrow 0$.
The presence of trapped regions is addressed in section 3.
The matching to  an exterior  spacetime  is summarized  in section 4.

\section{The scalar field model}

In order to study black hole formation in $AdS_4$ we consider a scalar field model coupled to gravity with negative cosmological constant, $\Lambda = - 3/l^2$, given by the action (in units $G=c=1$)
\be
S = \int \sqrt{-g}~ d^4 x \Big[\frac{1}{2}( R + \frac{6}{l^2})~-~ \frac{1}{2} (\partial_\mu
\phi {\partial^\mu} \phi) - V (\phi) \Big]\;,
\label{1}
\ee
where $l$ is the  curvature radius of $AdS_4$.
The scalar field $\phi$ is assumed to depend on 
time only, $\phi = \phi (t)$.
Therefore we choose as an ansatz   the flat Friedmann-Robertson-Walker  (FRW) metric inside the scalar matter written in the form

\be
ds^2 = -dt^2 + a^2(t) [dr^2 + r^2d \Omega_2^2]~,
\label{5}
\ee
where $a(t)$ is the scale factor. 
In this co-moving frame the energy density $\rho(t)$ and pressure $p(t)$ of the scalar field are simply
\bqa\label{3}
\rho(t) &=& \frac{1}{2} \dot{\phi}^2 + V(\phi),\nonumber
\\
p(t) &=& \frac{1}{2} \dot{\phi}^2 - V(\phi)\;,
\eqa
where a factor of $8\pi$ is absorbed into the definition of $\rho$ and $p$.
The scale factor $a(t)$ and the scalar field  $\phi(t)$ are determined from  Einstein's equations 
\be
R_{\mu v} =  T_{\mu v} - \frac{1}{2} g_{\mu v} T^\rho_\rho -
\frac{3}{l^2}~ g_{\mu v},
\label{6}
\ee
which in this case reduce to
\be
\dot{a}^2 = \frac{1}{3} \rho a^2 - \frac{a^2}{l^2},
\label{7}
\ee
and
\be 
\ddot{a} = - \frac{1}{6} (\rho + 3p) a - \frac{a}{l^2},
\label{8}
\ee
where $\dot{} = d/dt$.
The  Klein-Gordon equation is given by
\be
\frac{d}{dt} (a^3 \dot{\phi}) = - a^3 \frac{dV (\phi)}{d\phi}.
\label{9}
\ee
One can check that Eq.~(\ref{7}) is compatible with the ansatz
\be
\dot{a}^2 = a^{2 \beta} - \frac{a^2}{l^2},
\label{10}
\ee
which then gives
\be
\rho (t) = 3\, a^{-2(1-\beta)}, 
\label{11}
\ee
such that the energy density becomes a function of the scale factor $a(t)$ only. 
In Eqs. (\ref{10}) and (\ref{11}) a proportionality constant of dimension $(1/length)^2$
is fixed to be one. Furthermore it is assumed that $\rho$ is restricted to
be written as a function of $a(t)$ only and that it is ruled by a
power-law with a singular
behaviour in the limit $a \rightarrow 0$ for $\beta < 1$, which finally allows an
explicit and exact solution of Einstein's equations. The analogous
case without the cosmological constant $(l \rightarrow \infty)$ is treated in refs.
\cite{Giambo:2005se} and \cite{Goswami:2004ne, Goswami:2005fu}

The equation for the  scalar field, which  can be obtained from (\ref{7}) and (\ref{8}) by using  (\ref{3}) and the form of the density (\ref{11}), reads 
\be
 {\dot{\phi}}^2 = 2 (1 - \beta) {a^{2(\beta -1 )}}~.
\label{13n}
\ee
From this it follows that the potential can be expressed in terms of the time derivative of the scalar field 
\be\label{2}
V=(2+\beta)a^{2(\beta-1)}=\gamma \dot{\phi}^2\;,
\ee
where we introduced a new parameter 
\be\label{12}
 \gamma=\frac{2+\beta}{2(1-\beta)}\;.
\ee
Note that  we choose $\gamma \geq 0$ so that the potential is bounded from below. This also constrains $\beta$ to be in the interval  $-2 \leq \beta <1$. Later, we show that this potential is consistent with an exponential dependence on $\phi$ (see Eq.~(\ref{22})).
Using this form of the potential the  equation of state (EoS) obtained from (\ref{3})  is given by
\be
p = w\rho = \frac{1-2 \gamma}{1+2 \gamma} \rho, 
\label{4}
\ee
where $-1<w \leq 1$ because $\gamma$ is a non-negative constant  $0 \leq \gamma$.
We have introduced three parameters, $\gamma$, $w$ and $\beta$, which are related to each other, thus   the model depends only on one free parameter (besides the curvature radius $l$).
For $\beta=-1/2$, corresponding to $\gamma=1/2$,  the pressureless case  is obtained  as can be seen from (\ref{3}).
We will later see that the sign of $\beta$ is responsible for the formation of  an apparent horizon ($\beta<0$) or a naked singularity ($\beta \geq 0$).\\

We now give analytic solutions for $a(t)$ and $\rho(t)$.
Eq.~(\ref{10}) can be  integrated with the solution
for the collapse
\be
a (t) = \bigg\{ l \sin \Big[ (1-\beta) \frac{t_s - t}{l} \Big] \bigg\}^{\frac{1}
{1 - \beta}},
\label{15}
\ee
where we call the integration constant $t_s$, because the singularity occurs when the  scale factor vanishes  $a (t_s) = 0$.  Also note that the solution must  satisfy $\dot{a}<0$  in order to have a collapse. We choose the initial condition $a (t=0) = 1$ by convention. This implies that $l \ge 1$ and 
the energy density is initially the same, $\rho(t=0) =3$, for any $\beta$.
Furthermore, this initial condition gives the  characteristic singularity formation time 
\be
\frac{t_s}{l} = \frac{1}{1-\beta} \arcsin \lk l^{-1}\rk,
\label{16}
\ee
which decreases as $l$ increases and requires  $\beta<1$. 
It can be seen from  (\ref{11}) that  the density $\rho$ diverges as $a \rightarrow 0$.
In the time interval $0 \leq t \leq t_s$,
we have $\dot a \leq 0$ and $\dot{a} = 0$ only at $t = t_s$.
Therefore the scale factor monotonically decreases on the whole time interval, and there is no bouncing solution in this model.

\begin{figure*}
\centering
\includegraphics[width=9cm]{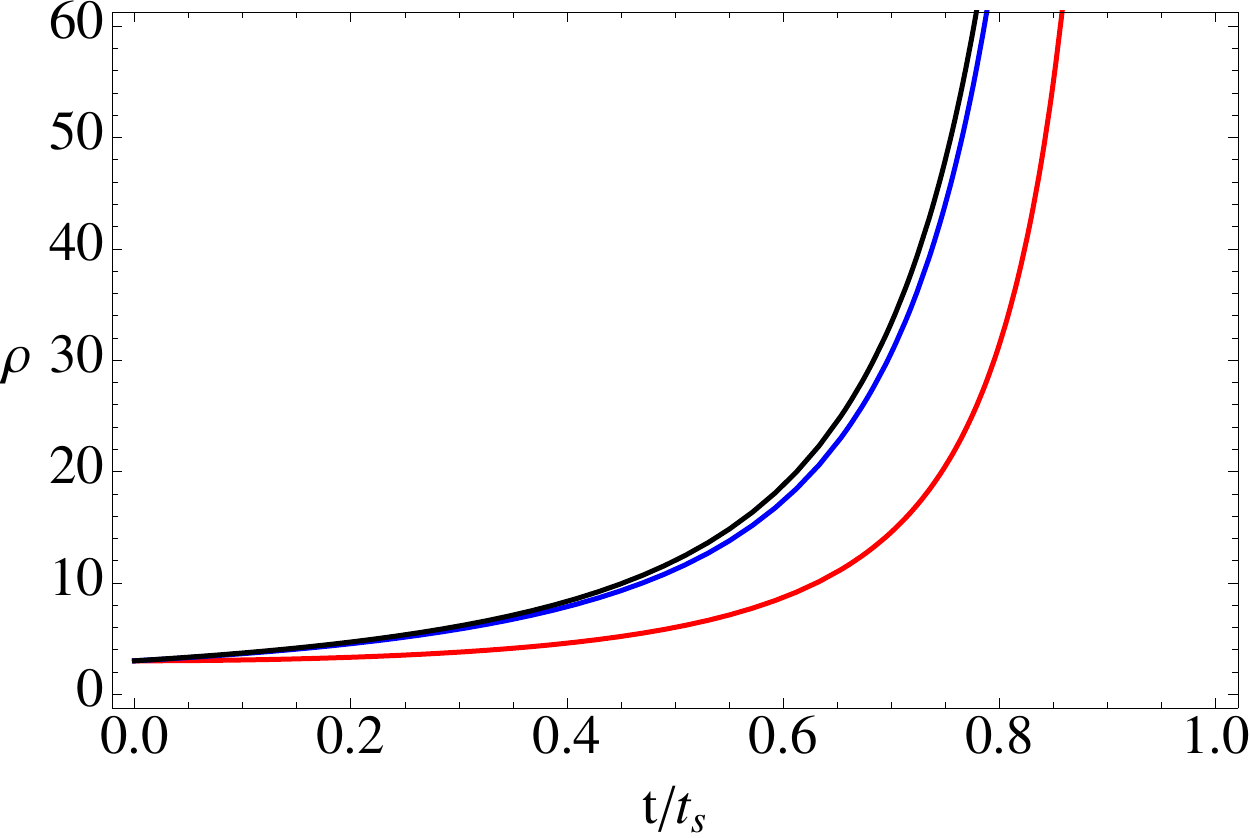}
\caption {The energy density as a function of the scaled time coordinate for different values  of the AdS curvature radius $l=1$ (red), $l=2$, (blue) and $l=\infty$ (black).
}
\label{density}
\end{figure*}

In terms of the scaled time variable $t/t_s$ one notes that
\be
a(t) = \bigg\{ l \sin \Big[\arcsin\lk l^{-1}\rk (1  - t/t_s) \Big] \bigg\}^{\frac{1}
{1 - \beta}}~
\label{15n}
\ee
 scales in $t/t_s$ for fixed parameters $\beta$ and $l$.
The limit of vanishing cosmological constant $l \to \infty$ (in the following denoted by subscript $\infty$) is given by
\be
a_{\infty} (t) = \bigg\{ (1-\beta)(t_s - t) \bigg\}^{\frac{1}{1-\beta}} =
\bigg\{ 1 -  t/t_s \bigg\}^{\frac{1}{1-\beta}},
\label{17}
\ee
since $t_s = \frac{1}{1-\beta}$ in the limit $l \to \infty$.
For $\beta = - \frac{1}{2}$ the solution (\ref{15}) coincides
with the Oppenheimer-Snyder (OS) model \cite{Oppenheimer:1939ue}, namely for
$OS-AdS_4$ as derived earlier in \cite{Ilha:1996tc}, and (\ref{17}) with the
one for OS $(k = 0)$ given in \cite{Giannoni:2004di,Giddings:2001ii}.
In \cite{Harada:2006dv} the case of the expanding flat Friedmann universe
with a scale factor $a \propto t^q, q = 1/(1-\beta)$ is discussed.
The solution for the energy density $\rho(t)$ can be obtained from Eqs.~(\ref{11}) and (\ref{15}):
\be
\rho (t) = \frac{3}{\Big\{ l \sin \Big[ (1-\beta) \frac{t_s-t}{l} \Big]
  \Big\}^2}~ = 
 \frac{3}{ \bigg\{ l \sin \Big[\arcsin[l^{-1}] (1  - t/t_s) \Big] \bigg\}^2}  ~,
\label{18}
\ee
which is plotted as a function of $t/t_s$ in Figure~{\ref{density}}.
The energy density $\rho(t)$ does not explicitly depend on
$\beta$, and for $l \to \infty$
\be
\rho_{\infty} = \frac{3}{(1-\beta)^2 (t_s -t)^2}~ =  \frac{3}{(1 -t/t_s)^2}~  ,
\label{19}
\ee
as  is known for the homogeneous OS model (see \cite{Adler:2005vn} Eq.~(58)).

We observe that near $t \simeq t_s$, the scale factor $a (t)$ and the energy density $\rho (t)$ are independent of the cosmological constant $\Lambda = - \frac{3}{l^2}$.
In the limit $t \rightarrow t_s$,
both $a(t)$ and $\rho(t)$ become $a_\infty$ and $\rho_\infty$, respectively,  not explicitly depending on $l$.
Furthermore, the behavior of the energy density near $t \simeq t_s$ is rather universal in this model because $\rho_\infty$ does not explicitly depend on the cosmological constant or on the EoS.\\

\begin{figure*}
\centering
\includegraphics[width=7.8cm]{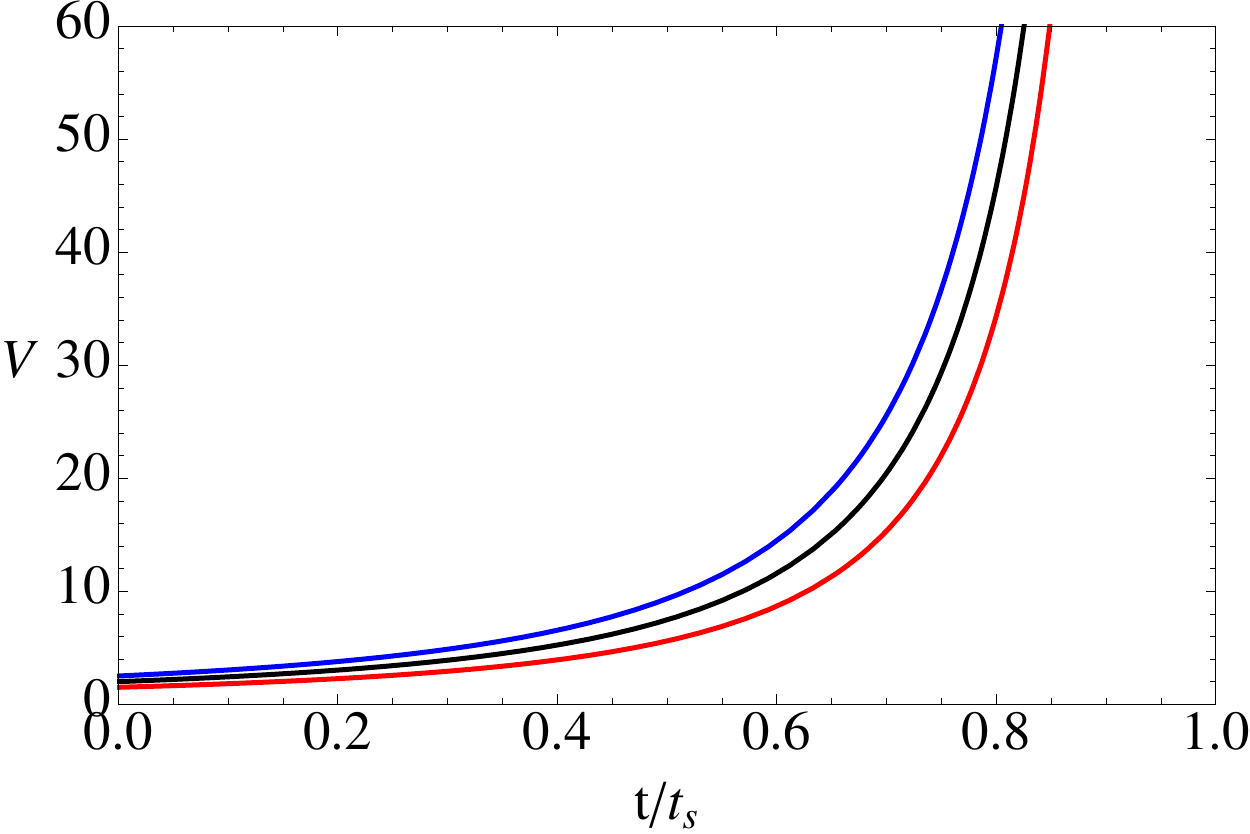}$\;\;\;\;\;\;\;\;$\includegraphics[width=7.8cm]{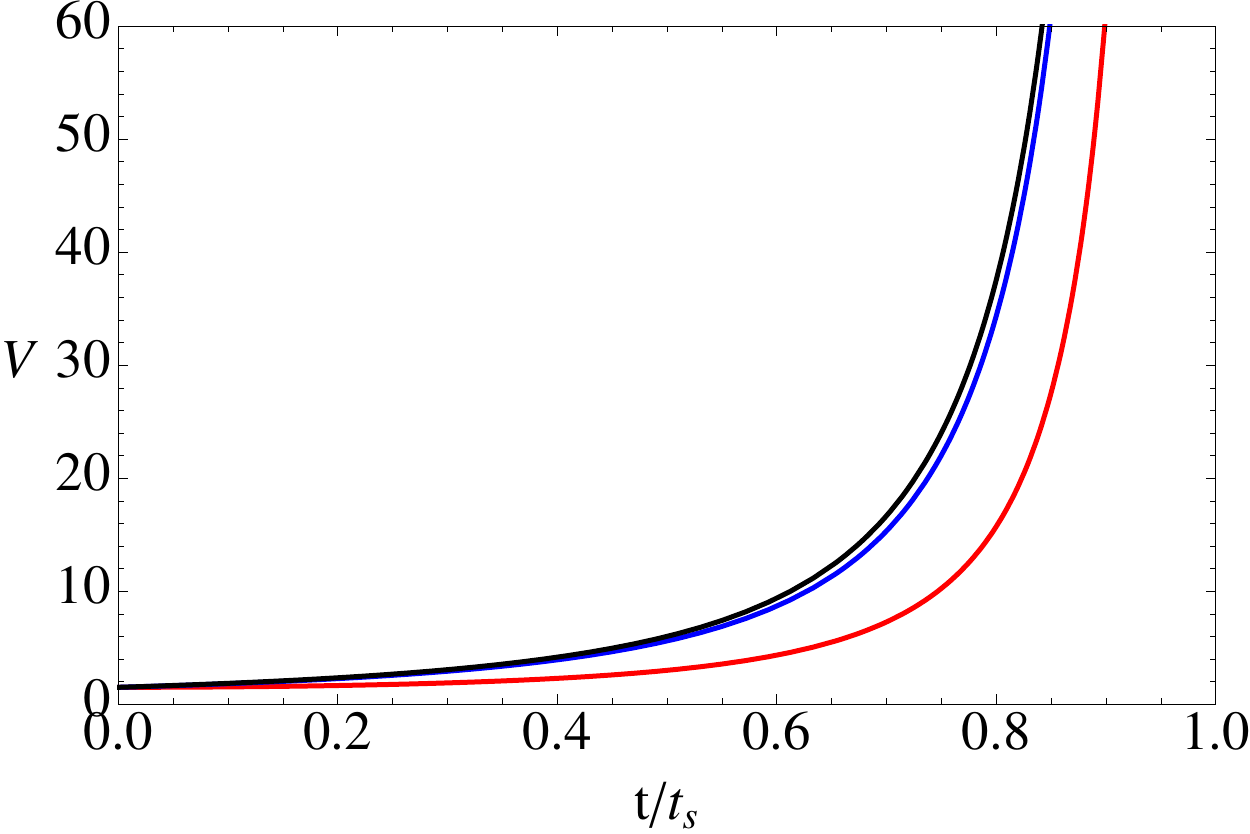}
\caption {Left: The potential for $l=2$ as a function of the scaling time for different values of $\beta=1/2$ (blue), $0$ (black), $ - 1/2$ (red).
Right: The potential for different values of $l=1$ (red), $l = 2$ (blue),
 $l = \infty$ (black) for $\beta=-1/2$.
}
\label{RR}
\end{figure*}

\begin{figure*}
\centering
\includegraphics[width=9cm]{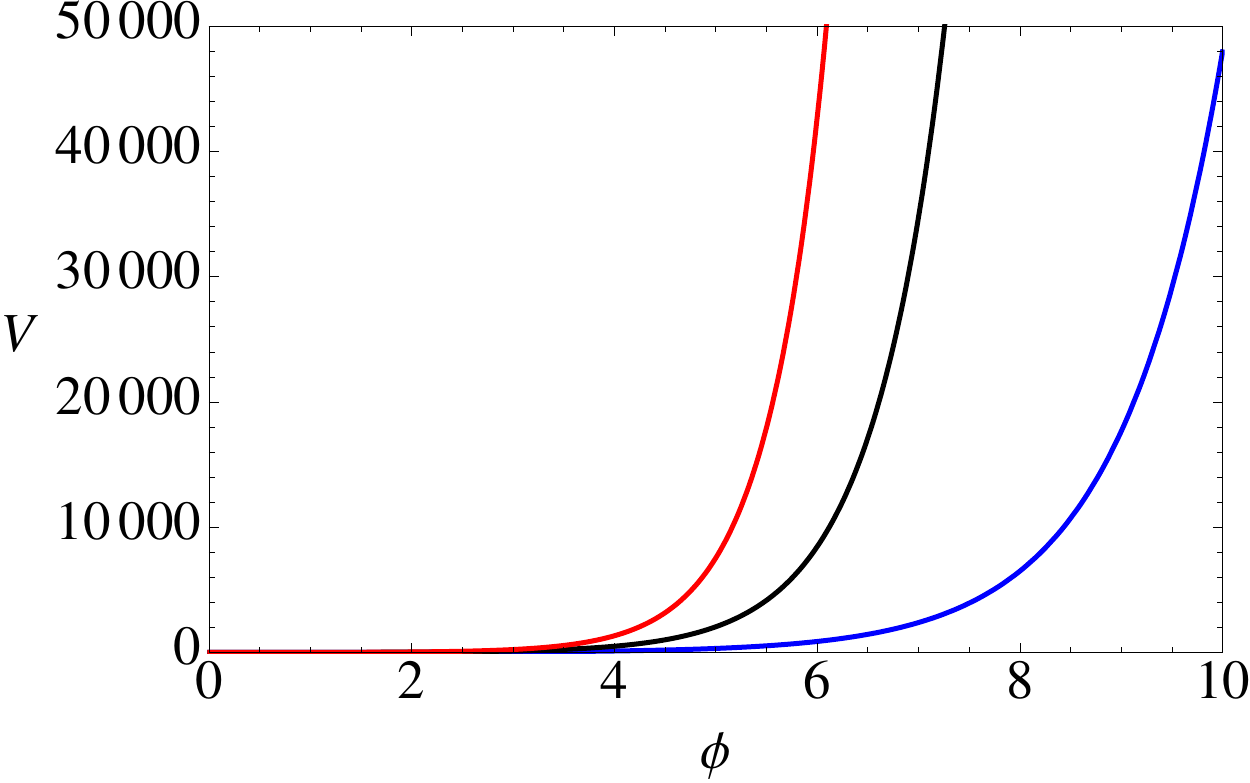}
\caption {The potential  as a function of the scalar field for $l=2$ and different values of $\beta=1/2$ (blue), $0$ (black), $ - 1/2$ (red).
}
\label{RR1}
\end{figure*}

The solution for the scalar field  $\phi$ can be obtained from (\ref{13n}).  Noting that $\dot{\phi} = \dot{a} \frac{d \phi}{da}$ and using (\ref{10}) one finds
\be
\frac{d \phi}{da} = \sqrt{2(1-\beta)} \frac{a^{\beta -1}}
{\sqrt{a^{2 \beta} - \frac{a^2}{l^2}}}.
\label{13}
\ee
One can check that this expression satisfies the Klein-Gordon equation (\ref{9})  by using
$\frac{dV}{d\phi} = \left( \frac{dV}{da} \right) \frac{da}{d\phi}$. The  equation  for the scalar field (\ref{13}) can be integrated to yield
\bqa
\phi &=& \sqrt{2(1- \beta)} \int^1_a \frac{a^{\beta -
    1}}{\sqrt{a^{2\beta} - \frac{a^2}{l^2}}}~ da \nonumber\\
    & =&\sqrt{\frac{2}{1-\beta}}  \log \left[ \frac{a^{\beta-1}+\sqrt{a^{2(\beta -1)} - 1/l^2}}{1+ \sqrt{1-1/ l^2}} \right] ,
\label{20}
\eqa
where we have chosen $\phi (t=0) = \phi (a=1) = 0$, using the translational symmetry of $\phi$, which follows from Eqs. (\ref{1}) and (\ref{2}). The scalar field $\phi$ diverges at the singularity formation time,  $t \to t_s$, for any   EoS. 
In the limit  $l \to \infty$ the above expression simplifies to 
\be
\phi_\infty = - \sqrt{2(1 -\beta)} \log a_\infty 
= -\sqrt{\frac{2}{1-\beta}} \log (1- t/t_s)
\;.
\label{21}
\ee
The potential (\ref{2}) is exponential in $\phi$, as can be seen  by inverting (\ref{20}) and using  (\ref{2}) 
\be
V= (2+\beta) \left[  \frac{1/ l^2 +\left(1+ \sqrt{1-1/l^2}  \right)^2 e^{2\,\sqrt{\frac{1-\beta}{2}} \phi}}{2  \left(1+ \sqrt{1-1/l^2}  \right) e^{\sqrt{\frac{1-\beta}{2}} \, \phi} }\right]^2\;\;,
\label{22}
\ee
which gives $V_\infty= (2+\beta) e^{\sqrt{2(1-\beta)} \phi_\infty } = (2+\beta)  (1-t/t_s)^{-2}$ in the limit $l\rightarrow \infty$ \cite{Goswami:2004ne} .
The potential is plotted as a function of the scaled time variable $t/t_s$ and the scalar field $\phi$  in Figure {\ref{RR}} and {\ref{RR1}}, respectively.

For $-~2< \beta < 1$ the potential diverges in the near-singularity region $t \simeq t_s$, because the potential becomes $V \simeq (2+\beta) (1- t/t_s)^{-2}$, and therefore it is not negligible in this region. 
Only when $\beta$ is at the lower bound, $\beta=-2$ ($\gamma=0$), then the potential is zero (the same EoS as for the stiff fluid). This special case corresponds to a model explored in Appendix B of
\cite{Hubeny:2004cn} (following the papers by Hertog et al.
\cite{Hertog:2003zs,Hertog:2003xg,Hertog:2004gz}), 
where cosmic censorship violation in AdS is discussed for a  potential that is negligible in the near singularity region.
In this respect the model under consideration differs significantly from the model of \cite{Hubeny:2004cn,Rangamani:2004iw} and allows naked singularity solutions for $\beta \geq 0$, as we  show in the next section. 
 \\
 
In {\cite{barrow}} an exact trigonometric solution to Einstein's equations
describing the evolution of inflationary universe models, driven by the
evolution of a scalar field is given. The self-interacting potential
contains a constant negative part acting like a cosmological constant.
In the notation of {\cite{barrow}} the two-parameter $(A,\lambda)$ solution
reads
\be
a(t) = a_0 \sin[ 2 \lambda t]^{A^2/2}, ~~~ a_0 = \rm{constant},
\ee
 describing gravitational collapse from $a _{max}$  to
$a = 0$ at $t_s = \pi/(2 \lambda)$. 
By comparison with (14) we identify
\be
t_s -t \rightarrow t ,~~~ A^2/2 = 1/(1-\beta)~~\rm{and}~~ 2\lambda= (1-\beta) /  \it{l}
\ee
Inserting $a(t)$ of (\ref{15}) into the equation (\ref{20}) of  the scalar field
one can rewrite 
\be
\phi = - \phi_B - \sqrt{2/(1-\beta)}\ln[l + \sqrt{l^2-1}],
\ee
where $\phi_B$ is the scalar field given in \cite{barrow}
\be
\phi_B = \sqrt{\frac{2}{1-\beta}} \,\ln\left[\tan\left[\frac{(1-\beta)t}{2l}\right]\right].
\ee
 With this field and adding the cosmological term $-3/l^2$
the potential given in \cite{barrow} is obtained
\be
V_B = \frac{2 + \beta}{l^2} \sinh^2\left[\phi_B\sqrt{\frac{1-\beta}{2}}\right] - \frac{1-\beta}{l^2}.
\ee
Besides the initial conditions, inflation versus collapse, the
trigonometric solution (Eqs. 16 - 20 in \cite{barrow}) agree with the expressions 
 presented here, i.e. Eqs. (\ref{15}), (\ref{20}) and (\ref{22}).

\section{Apparent horizon and trapped region}

In order to determine the dynamics of apparent horizons and the relation to black hole formation the definition of trapping horizons is crucial  \cite{Hayward:1994,Hayward:1996,Hayward:2000ca,Booth:2005qc,Booth:2005ng,Ashtekar:2004cn}. Next we will  analyse under which conditions  trapped surfaces form and investigate their properties.

\subsection{Marginally trapped surfaces}
For the FRW metric (\ref{5}) the following null geodesics are defined (see also \cite{Poisson:2004}),
\be
 l_{\mu} = ( -1, a(t), 0, 0)~, ~~n_{\mu} = \frac{1}{2} (-1 , -a(t), 0 ,0),  \qquad  
\label{tr1}
\ee
with $l_{\mu}l^{\mu} = n_{\mu} n^{\mu} = 0$ and  $l_{\mu} n^{\mu} = -1$.
Next the outgoing (ingoing) null expansions $\theta_+~(\theta_-)$ are introduced
\be
\theta_{+} = h_{\mu \nu} \nabla^{\mu} l^{\nu},~~~ \theta_{-} = h_{\mu \nu} \nabla^{\mu} n^{\nu},
\label{d+}
\ee
where  the transverse metric 
\be
h_{\mu \nu} = g_{\mu \nu} + l_{\mu} n_{\nu} + n_{\mu} l_{\nu}~,
\label{hmu}
\ee
satisfies $h_{\mu \nu} l^{\nu} = h_{\mu \nu} n^{\nu} =0$.

For the FRW metric (\ref{5}) and introducing $R(r,t) = r a(t)$, which corresponds to the physical radius of the collapsing matter, one obtains \cite{Booth:2005ng}
\be
\theta_+ = 2 \frac{\dot{R} + 1}{R} = 2 \(\frac {\dot{a}}{a} + \frac{1}{a\,r}\)~,~~~~\theta_-=  \frac{\dot{R} - 1}{R} =  \frac {\dot{a}}{a} -  \frac{1}{a\,r}~,
\label{th+}
\ee
which leads to the expansion
\be
\theta = \theta_+ \theta_- = \frac{2}{a^2} \( {\dot{a}}^2 - \frac{1}{r^2}\)~.
\label{thet}
\ee
Marginally trapped surfaces have the property  $\theta = 0$,
\be
 \theta_{+} = 0,~~~  \theta_{-} < 0~,
\label{trapp}
\ee
which gives the location of the apparent horizon 
\be
r_{AH}= - \frac{1}{ \dot{a}(t) }.
\label{cond}
\ee

An equivalent definition for the boundary of a possible apparent horizon is obtained from
\be
g^{\mu\nu} \partial_\mu R\, \partial_\nu R = 0,
\label{23}
\ee
which  for the metric at hand  (\ref{5}) becomes
\be
\dot{R}^2 (r,t) = r^2 \dot{a}^2 (t) = 1.
\label{24}
\ee
The above condition can be translated into a condition on the  Misner-Sharp mass \cite{Misner:1964je} given by 
\be
2 m(r,t) = R (1-g^{\mu \nu} \partial_\mu R \partial_\nu R) = r^3 a (t) \dot{a}^2
(t),
\label{32}
\ee
from which it follows that marginally trapped surfaces occur whenever 
\be
R(r,t) = 2m (r,t).
\label{31}
\ee
Using Eq.~(\ref{10}) one may rewrite Eq.~(\ref{31}) as
\be
{\dot{a}}^2 = a^{2\beta} - a^2/l^2 = 1/r^2~.
\ee
Assuming that the spherical ball of the collapsing scalar field is confined to a sufficiently  small radius $r_b$, i.e. $0 \le r \le r_b $, the above equation is only solved for  $\beta < 0$.
This shows  that an apparent horizon protecting the singularity is only formed if $\beta<0$,  implying $\gamma < 1$ and  $w=p/\rho > - 1/3$.
 
\begin{figure*}
\centering
\includegraphics[width=12cm]{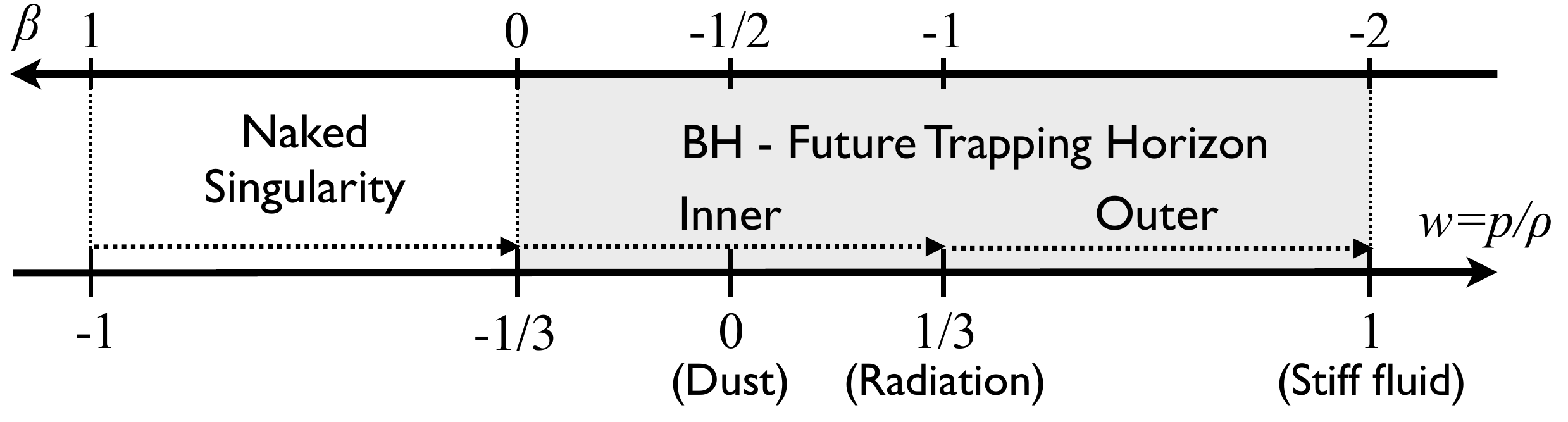}
\caption {Spherical scalar field collapse: physical conditions by EoS for black hole vs. naked singularity formation in $AdS_4$
(for $ r_b<1$).
}
\label{R5}
\end{figure*}

In this case the marginally trapped surface ($\theta_+=0$) can be further classified into a future inner/outer trapping horizon given by the condition \cite{Hayward:1994}
\bqa
\rm{future~ inner}&:&~~\theta_-<0~~\&~~\partial_- \theta_+>0~,\nonumber \\ 
\rm{future~ outer}&:&~~\theta_-<0~~\&~~\partial_- \theta_+<0~,
\eqa
where $\partial_-=2n^\mu\partial_\mu=(\partial_t-\frac{1}{a} \partial_r)$ is the Lie  derivative along future directed ingoing null geodesics.
With the help of  Einstein's equations  (\ref{7}) and (\ref{8}) and $\theta_+=0$ the condition becomes
\be
2 n^{\mu} \partial_{\mu} \theta_+ = 
(1 + \beta) \( \frac{1}{r^2 a^2 } - \frac{1}{l^2}\)~,
\label{inout}
\ee
which is positive (future inner horizon)  for $ - 1 < \beta < 0$ and negative (future outer horizon)  for $\beta<-1$ in the  near the singularity region where $r_b< l$.  The different possibilities are summarized in Figure {\ref{R5}}  
in terms of the EoS parameter  $w=p/\rho$ and $\beta$.\\

The apparent horizon condition (\ref{cond}) for  the solution (\ref{15}) reads
\be
\frac{1}{r_{AH}} = \cos \Big[ (1-\beta) \frac{t_s-t}{l}
\Big] \Big\{ l \sin \Big[(1-\beta) \frac{t_s - t}{l} \Big]
\Big\}^{\frac{\beta}{1 - \beta}}.
\label{25}
\ee
It is clear from this equation that for any value of $l$, the apparent horizon can only   reach  the singularity at $t=t_s$ if and only if $\beta$ is negative. Therefore, the sign of $\beta$ is the only factor that  determines the formation of a black hole in this model. On the other hand, when $\beta$ is positive, there is no trapped region, and thus a naked  singularity  can form.

Assuming in the following a sufficiently small  matching radius  (i.e. the boundary of the collapsing scalar field) $r = r_b<1 $, one observes from (\ref{25}), that for $\beta = 0$ (and positive $r_{AH}$),
\be
r_{AH} = \frac{1}{\cos \Big[ (t_s - t)/l \Big]} \ge 1~,
\label{26}
\ee
such that the horizon is outside the physical region, that is, outside the
boundary surface $r = r_b$. From Eq.~(\ref{25}), one can show that $r_{AH}$ is always outside the physical region for $\beta \geq 0$.

For $\beta \ge 0$ no trapped region is formed, instead a naked
singularity is present without being shielded  by a horizon.
The same behaviour is found and  discussed in \cite{Goswami:2004ne} for $l \to \infty$, where
(\ref{25}) becomes
\be
\frac{1}{r_{AH}} = \Big\{ (1-\beta) (t_s - t) \Big\}^
{\frac{\beta}{1 - \beta}} \,,
\label{28}
\ee
and for $\beta =  \frac{1}{2}$, where $t_s = 2$,
\be
r_{AH} = \frac{2}{t_s - t},
\nonumber
\ee
compared with $\beta = 0$, which gives $r_{AH} = 1 > r_b$.
The physical ``interior'' region $r \le r_b$ is not trapped for
$\beta \ge 0$. This statement is in contradiction with the claim in  \cite{Ganguly:2012xr},
which allows the formation of an apparent horizon even for $\beta \ge 0$.

For negative $\beta,~\beta < 0$, the behaviour of the solution $a(t)$
(\ref{15})  and consequently Eq.~(\ref{25}) is different: for any shell of
the scalar field with $r \le r_b$ an apparent horizon is formed during
the time evolution of the system (\ref{1}). As a representative example
consider $\beta = - \frac{1}{2}$ with 
\be
a(t) = \Big\{ l \sin \Big[ \frac{3 (t_s - t)}{2 l} \Big] 
  \Big\}^{\frac{2}{3}}  ,
\label{29}
\ee
which is equivalent to the Oppenheimer-Snyder model with a negative cosmological constant as derived in \cite{Ilha:1996tc}.
In this case  the horizon boundary is located at 
\be
\frac{1}{r_{AH}} = \cos \Big[ \frac{3}{2}~ \frac{t_s - t}{l} \Big]
\Big\{ l \sin \Big[ \frac{3}{2}~ \frac{t_s - t}{l} \Big] \Big\}^{-
  \frac{1}{3}},
\label{30}
\ee
with $t_s = \frac{2}{3} l \arcsin [l^{-1}]$.

\subsection{Conformal time analysis}
A convenient and transparent way to present the
different properties of the solution depending on the sign of the parameter $\beta$ is obtained by introducing the conformal time $\eta$,
\be
\eta= \int^t_0 \frac{dt'}{a(t')} =\int^{a(t)}_{a(0)} \frac{da}{a \,\dot{a}}.
\ee
Using (\ref{10}) this can be integrated to give in terms of hypergeometric functions
\bqa
\eta&=&\frac{a^{-\beta}}{\beta} \left._2 F \right._1\lk\frac{1}{2},  \frac{- \beta}{2(1- \beta)}; 1-\frac{\beta}{2(1- \beta)}; \frac{a^{2(1- \beta)}}{l^2}\rk \nonumber\\
&~-&
\frac{1}{\beta} \left._2 F \right._1\lk\frac{1}{2},  \frac{- \beta}{2(1- \beta)}; 1-\frac{\beta}{2(1- \beta)}; \frac{1}{l^2}\rk\;.
\eqa
As already noted the behaviour of $a(t)$ near $t \simeq t_s$ does not 
explicitly depend on $l$ and therefore its sufficient to only look at the $l\to \infty $ limit where 

\be
\eta_\infty  = - \frac{1}{\beta} \left(1 -
\frac{1}{a^\beta_{\infty}}\right). 
\label{33}
\ee
The apparent horizon formed during the collapse of the scalar field sphere $r\le r_b$ has the boundary at  $a^\beta_{AH} = \frac{1}{r_{AH}}$ leading to 
\be
\eta_{AH} = - \frac{1}{\beta} (1 - r_{AH}).
\label{34}
\ee

\begin{figure*}
\centering
\includegraphics[width=9cm]{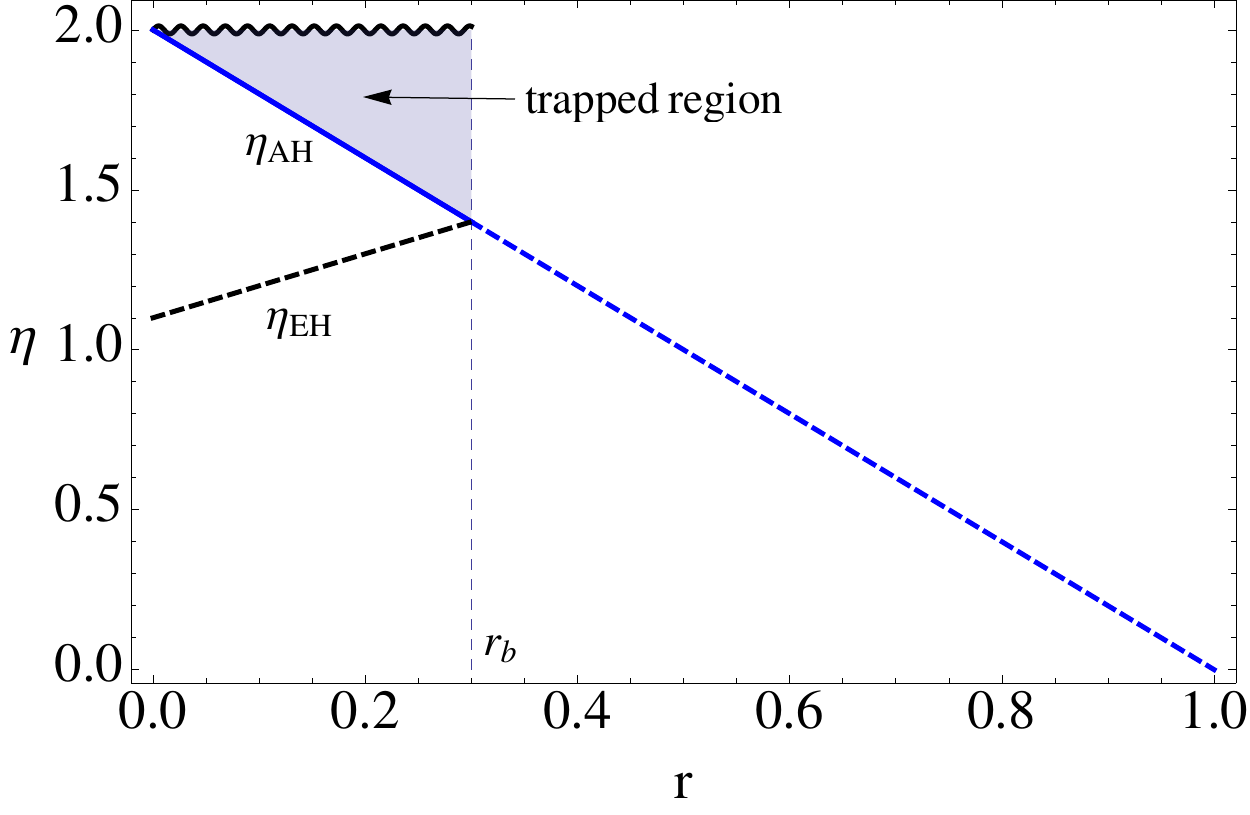}
\caption {Event and apparent horizon shown in the conformal time  for $\beta=-1/2$ in the $l=\infty$ limit. The shaded area represents  the trapped region and the singularity of zero radius occurs at the conformal time  $\eta = \eta(a(t_s)) =2$. The region $r>r_b$ is unphysical.}
\label{R4}
\end{figure*}
The inner event horizon is determined by a radial light ray emitted at $r=0$ which just reaches the surface at $r=r_b$ at the same position as the apparent horizon \footnote[1]{
When the inner spacetime is matched to an outer region the event horizon stays at the surface $r=2M$ once it intersects the apparent horizon. This is done in section 4 and depicted in Fig.  \ref{Adler}.
} 
\be
\eta_{EH} = 
r + \frac{1}{\vert \beta \vert} (1 - r_b) - r_b~.
\label{35}
\ee
This is shown in Fig.~\ref{R4} for the black hole case for $\beta=-1/2$. Light emitted at $\eta > \eta_{EH}$
(at $\eta < \eta_{EH}$) ends up at the singularity (reaches an observer at
$r \to \infty$) \cite{Hartle}.
For positive $\beta,~ \beta > 0$, the situation is very different, namely
$\eta_{``AH''}$ $< 0 ~~ (\rm{for }~r  \le r_b)$ and there is no trapped
region formed in the physical region of the collapse.

Using the conformal time, we can identify a connection between the inner (outer) future trapped horizon and the time (space) like nature of the horizon.
For finite $l$ one derives, using (\ref{10}) and the condition (\ref{cond}),
\be
\frac{\partial r_{AH}}{\partial\eta} = a \,  \frac{\partial r_{AH}}{\partial t} = \beta - (1-\beta) r_{AH}^2 \frac{a^2}{l^2},
 \ee
 which in the near singularity region,  where $r_{AH}/l \ll1$, is proportional to
 \be
\left. \frac{\partial r_{AH}}{\partial\eta} \right|_{t\to t_s} \propto \beta.
 \ee
The apparent horizon is time-like for
$-1<\beta <0$ (``inner")
or space-like for $-2<\beta <-1$ (``outer"), c.f.~with (\ref{inout}).
For a related discussion in the context of an apparent cosmological horizon,
see Ref. \cite{Harada:2006dv}, especially Fig. 6.

\subsection{Energy conditions}
Next we investigate for which values of $\beta$ the weak and strong energy conditions hold.
The weak requirement
\cite{Joshi:2008zz,Poisson:2004} holds in terms of the energy density and pressure when 
\be
\rho \ge 0\;, \qquad \rho + p > 0\;,
\ee
which is satisfied for $\gamma\ge -1/2$, i.e. for both black hole and naked singularity formation.

The strong energy condition requires to make
\be
R_{\mu \nu} u^\mu u^\nu \ge 0
\label{str1}
\ee
 for every timelike vector $u^\mu$
\cite{Penrose:1969pc,Hawking} (see~{\cite{Poisson:2004}} for $l \to \infty)$.
Using (\ref{3}) and (\ref{6})
the condition becomes 
\be
\frac{1}{2}(\rho + 3 p) = (1 - \gamma) \dot{\phi}^2 \ge -3/l^2~,
\label{str2}
\ee
or
\be
\beta~ a^{2(\beta -1)} \le 1/l^2~.
\label{str3}
\ee
As $a \rightarrow 0$, the left-hand side goes to positive infinity if $\beta$ is positive. Therefore the strong condition  excludes the case of naked singularity formation.

\section{Matching with an exterior space}

In order to obtain the full spacetime the metric inside the collapsing scalar field has to be matched with the outer region. 
We first perform the matching by using the Schwarzschild-$AdS_4$ metric and then transform it to the Painleve-Gullstrand coordinate system to point out the fluid analogy.

\subsection{Schwarzschild-AdS coordinates}

Outside the collapsing  scalar field matter we choose the Schwarzschild-$AdS_4$ metric
\be
ds^2 = -f (Y) dT^2 + \frac{dY^2}{f(Y)} + Y^2 d \Omega_2^2, ~~f(Y) = 1 - \frac{2M}{Y} + \frac{Y^2}{l^2},
\label{37}
\ee
which has to be matched with the interior FRW metric (\ref{5}) at the boundary of a spherical junction hypersurface $r=r_b$. 
 From the angular parts of the two metrics, we can identify at the boundary 
\be
Y = r_b a(t) \equiv Y_b(t),
\label{40}
\ee
where $t$ is the proper time of the collapsing scalar field and $0 \leq Y_b \leq r_b$.
From the Israel junction conditions, the Misner-Sharp mass (\ref{32}) at the boundary is $2 m(r_b, t) =Y_b(t)(1-f(Y_b (t)))$, 
which then gives \cite{Giambo:2005se}
\be
2M (t) =\frac{Y^{2\beta+1}_b (t)}{r_b^{2(\beta-1)}},
\label{36}
\ee
leading to a time-dependent mass $M(t)$.   Note that   $2M (t = 0) = r^3_b$.
In addition the outside time coordinate must satisfy \cite{Giambo:2005se}
\be\label{mt}
\frac{dT}{dt} = \frac{1}{f(Y_b (t))} \equiv \frac{d  }{dt} T_b(t)
\ee
at the boundary and we set $T_b(0)=0$.  

As in the previous section the different cases $\beta \lessgtr 0$ are
to be considered. For $\beta = -1/2$, the OS type model, the outside metric becomes
\be
f(Y) = 1 - \frac{2M (t=0)}{Y} + \frac{Y^2}{l^2},
\label{42}
\ee
with an event horizon at $0<Y< r_b$.

For $\beta \geq 0$ (i.e., the naked singularity case), we observe that the metric smoothly transforms to vacuum AdS in the near singularity region, $t \rightarrow t_s$, because $2M(t_s)/Y$ is zero everywhere even at $Y=0$. For vanishing cosmological constant we would have Minkowski space instead as discussed in \cite{Goswami:2004ne}. 
The singularity formation time in  the outside   coordinate is given by
\be
T_s
= \int_0^{t_s} \frac{dt}{f(Y_b)} 
= \int_0^{t_s} \frac{dt}{1-r_b^2 a^{2 \beta}(t) + r_b^2 a^2(t)/l^2}\;.    
\ee   
Because $f(Y_b)$ with $r_b <1$ is positive ($0<f(Y_b) \leq 1$), $T_b$ is an increasing function of $t$ and $T_s$ is  positive  for $0 \leq \beta$ and for any curvature  radius ($1 \leq l$). 
The boundary of the scalar field collapses to a naked singularity, in accordance with Theorem 5.2 stated in \cite{Giambo:2005se}.
The function $f(Y)$ is bounded away from zero in the neighbourhood of the singularity at
$Y=0$ and $T = T_s$, and there exists a radial null geodesic that can escape from the singularity.

\subsection{Painleve-Gullstrand coordinates}
Following the treatment of gravitational collapse of uniform
perfect fluids as described in \cite{Adler:2005vn}, the Painleve-Gullstrand coordinate system \cite{Poisson:2004} is well suited  for  the scalar field collapse under consideration.
For the interior spacetime  the radial coordinate
\be
R = r\, a(t)
\label{A1}
\ee
is introduced, leading from (\ref{5}) to
\be
ds^2 = - [ 1 - \psi^2 (R,t)] dt^2 + dR^2 - 2 \psi (R,t) dRdt + R^2
d \Omega_2^2,
\label{A2}
\ee
where
\be
\psi (R,t) =  R \frac{\dot{a}}{a} = - \frac{R}{l} \cot
\Big[ (1 - \beta) \frac{t_s-t}{l} \Big].
\label{A3}
\ee
We used the fact that $\dot{a} < 0$, so we have $\psi < 0$.
For $l \to \infty$,
\be
\psi \to  \frac{-R}{1 - (1-\beta)t}~,
\label{A4}
\ee
the collapse to ``zerosize''  occurs at $t \to t_s = \frac{1}{1 - \beta}$.

The exterior region with the metric (\ref{37})  can be brought into the form of Painleve-Gullstrand (\ref{A2}) by  the transformation
\be
Y=R ~~\mbox{and}~~ T = t + g (R),
\label{A5}
\ee
where 
\be
\frac{dg}{dR} =  \frac{\psi}{1 - \psi^2}, ~~\mbox{with}~~\psi^2 = \frac{2M (t)}{R} - \frac{R^2}{l^2}.
\label{A7}
\ee
We match $\psi$ at the boundary to get the matching condition:
\be
\frac{1}{l} \cot \Big[ (1 - \beta) \frac{t_s - t}{l} \Big] =
\sqrt{\frac{2M(t)}{R^3_b} - \frac{1}{l^2}},
\label{A8}
\ee
where $R_b $ denotes  the boundary of the fluid sphere. 
As discussed already, in the near singularity region the system is dominated by the limiting 
 behaviour for $l \to \infty$,
especially for small spheres $r_b < 1$. 
This limit is transparently treated by two representative examples.

\begin{figure*}
\centering
\includegraphics[width=9cm]{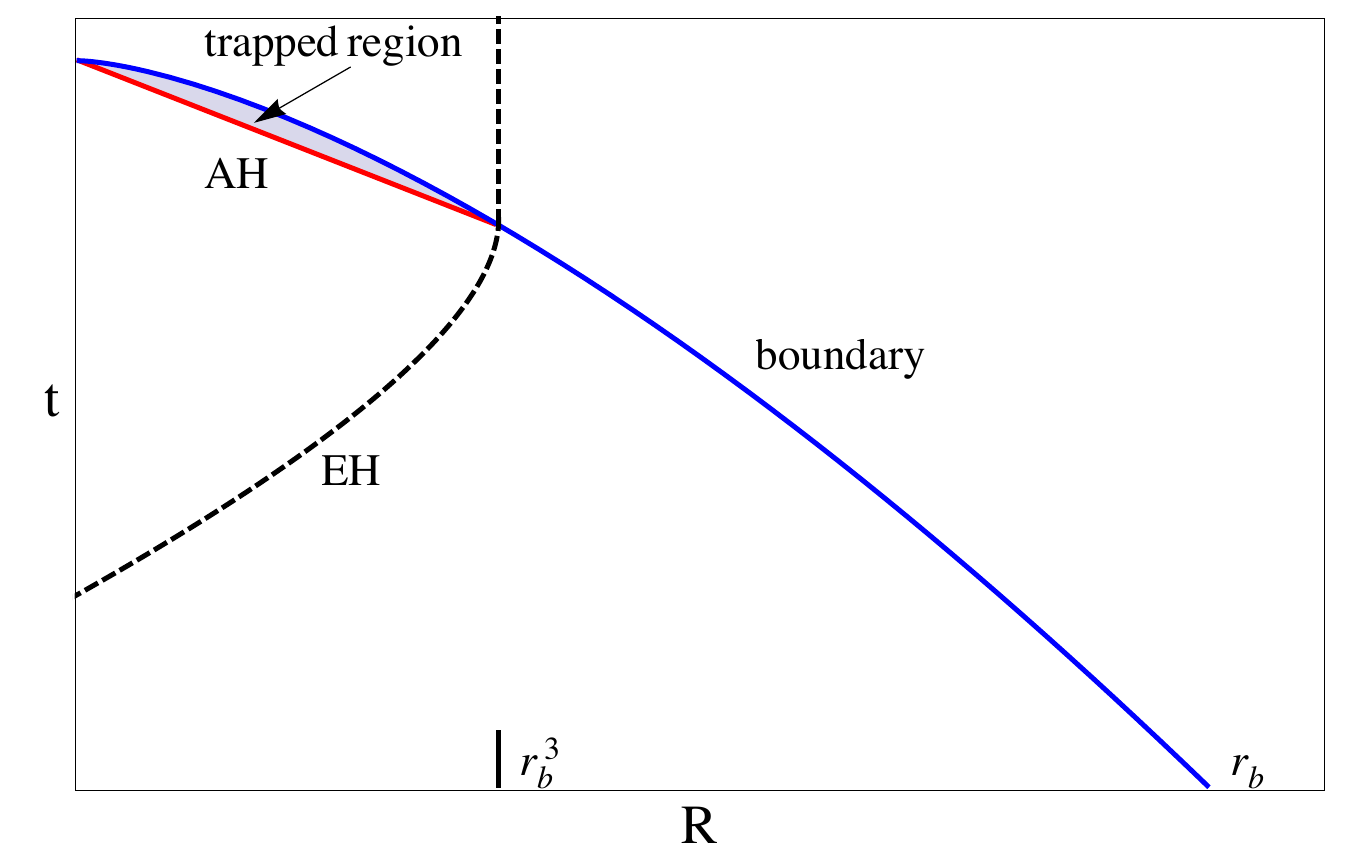}
\caption{Schematic diagram for black hole formation. Apparent horizon (AH) and event horizon (EH) are plotted together
with the boundary of the scalar field sphere (fluid).}
\label{Adler}
\end{figure*}

In the black hole formation case  $\beta<0$, we choose $\beta=-1/2$ for concreteness. The trajectory of the boundary of the collapsing matter is obtained from the matching condition (\ref{A8}), and the apparent horizon can be easily obtained from the infinite redshift, $g_{tt}=0$, which implies $\psi=-1$. For $l \rightarrow \infty$, we have the simple expressions
\bqa
&& t = \frac{2}{3} \left( 1 - \sqrt{\frac{R^3_b}{r^3_b}}\right), \nonumber \\
&& t = \frac{2}{3} (1-R_{AH}),
\label{A9}
\eqa
where $R_{AH}$ is the location of the apparent horizon. 
The event horizon in the inner region is given by (\ref{35}),
to be translated into
\be
\left. R(t) \right|_{EH} = (1 - 3t/2)^{2/3} [ 3 r_b - 2(1 - 3t/2)^{1/3}]~.
\label{A10}
\ee 
 The evolution of the spacetime is shown in Figure {\ref{Adler}} (c.f. Fig. 6 in \cite{Adler:2005vn}).
 The event horizon starts from  zero radius and then grows until it coincides with the fluid surface at $R=2M$. At this time the apparent horizon forms and shrinks to zero size  as the boundary of the collapsing matter reaches the singularity. The trapped region lies between the boundary and the apparent horizon.


The situation for positive $\beta$ is different. For $\beta = + \frac{1}{2}$ and $l \to \infty$, we have
%
\bqa
&& t = 2 \left( 1 - \sqrt{
\frac{R_b}{r_b}}\right), \nonumber \\
&& t = 2 (1 - R_{``AH"})~.
\label{A11}
\eqa
There is no trapped region inside the sphere of the collapsing scalar field,
i.e. a naked singularity is formed. This also holds for the AdS case $(l < \infty)$.

\section{Conclusions}

In this paper we studied the collapse of homogeneous spherically symmetric scalar field matter within a sufficiently small sphere ($r\le r_b$) in the presence of a negative cosmological constant $\Lambda = - 3 / l^2$. 
The model is rather independent of initial conditions, but depends only on one EoS parameter $\beta$ (equivalently $w$ or $\gamma$) which  decides the fate of the spacetime.
Either a naked singularity forms or the singularity is protected by a trapped surface resulting in black hole formation.
We showed that for $0 \leq \beta < 1$ but for any given value of the negative cosmological constant, there is generic formation of naked singularities as the final state of collapse, such that there is  cosmic censorship violation in anti-de Sitter space (first discussed in 
\cite{Hertog:2003zs,Hertog:2003xg,Hertog:2004gz} and debated in
 \cite{Hubeny:2004cn,Rangamani:2004iw}). 
Cosmic censorship violation was also discussed earlier in  similar models but without cosmological constant  \cite{Goswami:2004ne,Ganguly:2012xr,Goswami:2012pv}. Observational consequences by gravitational lensing to distinguish
(Schwarzschild) black holes from naked singularities to test cosmic
censorship
are discussed in \cite{Ellis}.

We investigated in detail how the formation of trapped surfaces depends on this parameter.
For $\beta<0$ an apparent horizon forms whereas for $\beta \geq 0$ the horizon lies outside the physical region and therefore the collapse terminates  in a naked singularity. 
The same result for vanishing cosmological constant has been found in \cite{Giambo:2005se} and \cite{Goswami:2004ne}. In this work we have shown that the presence of a cosmological constant does not affect this picture. 
A justification of this fact is provided by the observation that  near the singularity formation time  $t_s$ the scale factor and the energy density diverge and do not explicitly depend on the value of the cosmological constant as discussed below  equation (\ref{19}). Also the scalar field  and the potential show the same behaviour near the singularity formation time. 
In addition the energy density shows universal behaviour: it does not depend on the free parameter $\beta$ (see Eq.~(\ref{18})).

We found that the weak energy condition always holds in this model but the strong energy condition only holds for the case of black-hole formation and is violated for the case of naked-singularity formation.

In the case of black hole formation we showed that for $-1 \leq \beta<0$ we have a marginally trapped future inner horizon (i.e. space-like horizon), whereas for $-2 \leq \beta < -1$ we have a marginally trapped future outer horizon (i.e. time-like horizon).

\subsection*{Acknolwedgements}
The authors would like to thank C. Ecker, D. Grumiller, V. Keranen and O. Taanila for useful discussion 
 and P.~S.~Joshi for correspondence and comments.
 We thank J.~D. Barrow for informing us about his
paper {\cite{barrow}}.
HN was supported by the Sofja Kovalevskaja program of the Alexander von Humboldt Foundation and the Bielefeld Young Researcher's Fund. SS was supported  by the project  P-26328 of the Austrian Science Fund (FWF).

\section*{References}
\bibliographystyle{unsrt} 

\end{document}